\documentclass[pra,aps,twocolumn,showpacs,floatfix]{revtex4-1}
\usepackage{amssymb}

\usepackage{amsthm}
\usepackage{dcolumn}
\usepackage{bm}
\usepackage{graphicx}
\usepackage{amsmath}
\usepackage{color}
\usepackage{lipsum}

\usepackage[colorlinks,
            linkcolor=blue,
            anchorcolor=blue,
            citecolor=blue
            ]{hyperref}

\begin{document}

\title{Entangling Nuclear Spins by Dissipation in a Solid-state System}

\author{Xin Wang}\thanks{These authors contributed equally to this work.}
\author{Huili Zhang}\thanks{These authors contributed equally to this work.}
\author{Wengang Zhang}
\author{Xiaolong Ouyang}\author{Xianzhi Huang}\author{Yefei Yu}\author{Yanqing Liu}\author{Xiuying Chang}\author{Dong-ling Deng}\email{dldeng@tsinghua.edu.cn}\author{Luming Duan}
\email{lmduan@tsinghua.edu.cn}
\affiliation{Center for Quantum Information, IIIS, Tsinghua University, Beijing 100084, PR China}

\begin{abstract}
Entanglement is a fascinating feature of quantum mechanics and a key ingredient in most quantum information processing tasks. Yet the generation of entanglement is usually hampered by undesired dissipation owing to the  inevitable coupling of a system with its environment. Here, we report an experiment on how to entangle two $^{13}$C nuclear spins via engineered dissipation in a nitrogen-vacancy system. We utilize the  electron spin as an ancilla, and combine unitary processes together with optical pumping of the ancilla to implement the engineered dissipation and deterministically produce an entangled state of the two nuclear spins, independent of their initial states. Our experiment demonstrates the power of engineered dissipation as a tool for generation of multi-qubit entanglement in solid-state systems.
\end{abstract}

\maketitle

\newtheorem{corollary}{Corollary} \newtheorem{definition}{Definition} %
\newtheorem{example}{Example} \newtheorem{lemma}{Lemma} %
\newtheorem{proposition}{Proposition} \newtheorem{theorem}{Theorem} %
\newtheorem{fact}{Fact} \newtheorem{property}{Property} 

\begin{figure}[tbp]
\includegraphics[width=0.47\textwidth]{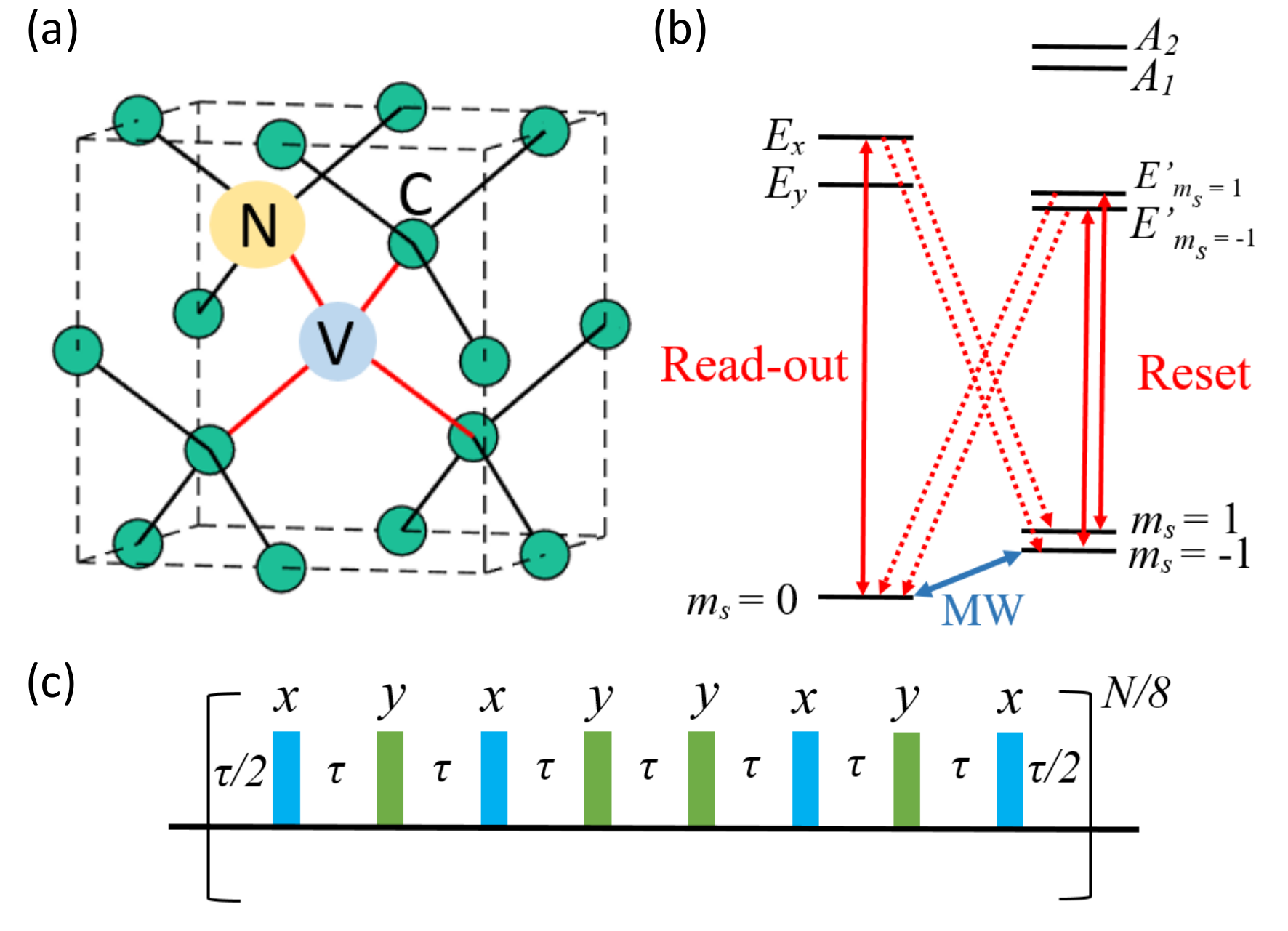}
\caption{The nitrogen-vacancy (NV) experimental system.
(a) The NV center structure in the diamond lattice, where the yellow circle represents the nitrogen atom, the blue circle represents the vacancy of a single atom, and green circles represent the carbon atoms. The red lines together with four atoms and a vacancy represent a NV center.
(b) The electron energy structure of the NV center. Two red lasers with wavelength of about 637.2nm are used in our experiment for read-out and reset of the electron spin state.
(c) The Carr-Purcell-Meiboom-Gill (CPMG) sequence for manipulating the electron and nuclear spins \cite{CPMG}.  This kind of XY-8 dynamical decoupling sequence is used to decouple different $^{13}$Cs from electron spin signals, thus making operating single and multiple nuclear spins possible.}
\label{NVIllustration}
\end{figure}

Entangled states are a crucial resource for various quantum technologies, ranging from quantum cryptography \cite{Gisin2002Quantum}, quantum metrology \cite{Giovannetti2011Advances}, to quantum computing \cite{Nielsen2010quantum}. Yet, faithful and reliable preparations of entangled states are usually hampered by decoherence and dissipation owing to the inevitable coupling  of a system with its environment. Traditionally, dissipation has been regarded as a detrimental factor since it would destroy unitary dynamics and wash out the desired entanglement.  A variety of notable approaches have been proposed to combat dissipation, including quantum error correction \cite{Gottesman2010Introduction}, topological qubits \cite{Nayak2008NonAbelian}, decoherence-free subspaces \cite{Duan1997Preserving,Zanardi1997Noiseless,Lidar1998Decoherence}, etc.  On the other hand, however, it has  been suggested that engineered dissipation can also be beneficial in preparing entangled states from arbitrary initial states \cite{Diehl2008Quantum,Cho2011Optical,Krauter2011Entanglement}. In particular, universal quantum computation is possible using only dissipation ~\cite{Cirac2009}. Experimentally, dissipative preparation of entanglement has been demonstrated with atomic ensembles \cite{Krauter2011Entanglement},  optical cavities \cite{Kastoryano2011Dissipative}, trapped ions \cite{Lin2013Dissipative,Barreiro2011Open}, and superconducting qubits \cite{Shankar2013Autonomously}.  Here, we extend this approach to the nitrogen-vacancy (NV) center systems (see Fig. \ref{NVIllustration} for a schematic illustration) and demonstrate the  preparation of entangled states for two $^{13}$C nuclear spins through engineered dissipation invoked by manipulation of the electron spin.

NV centers in diamond have emerged as one of the most promising experimental platforms for quantum information processing \cite{Jelezko2006Single,Dobrovitski2013Quantum,Doherty2013Nitrogen,Acosta2013Nitrogen,
Childress2013Diamond,yuan2017observation} and quantum sensing \cite{Degen2017Quantum}. They exhibit atom-like properties (such as long-lived spin quantum states and well-defined optical transitions) in a robust solid-state device,  with spin degrees-of-freedom coming from both their bound electrons and nearby nuclear spins. These spin states have a long coherence time even at room temperature and can be initialized and read out by lasers and manipulated by microwave pulses. In experiment, notable progresses have been made in demonstrating universal quantum gates \cite{Jelezko2004Observation,Sar2012Decoherence,rong2015experimental,Casanova2016Noise,zu2014experimental}, multipartite entanglement \cite{neumann2008multipartite,Bradley2019ATen,vanDam2019Multipartite},  quantum registers \cite{dutt2007quantum,jiang2009repetitive},  quantum error correction \cite{Waldherr2014Quantum,taminiau2014universal}, multipartite entanglement, entanglement distillation \cite{kalb2017entanglement}, quantum simulation \cite{yuan2017observation}, and quantum algorithms \cite{Sar2012Decoherence,Shi2010Room}, etc. Yet, hitherto \textit {no} experimental demonstration of how to prepare entangled states via engineering dissipation in NV systems has been reported to the best of our knowledge.  Such a demonstration would  offer novel prospects for open-system quantum information processing and simulation with NV centers.

In this paper, we add this missing block by experimentally demonstrating the deterministical preparation of  the maximally entangled two-qubit Greenberger-Horne-Zeilinger (GHZ) state through engineered dissipation with a NV center in diamond. More precisely, we prepare two $^{13}$C nuclear spins into the GHZ state through engineered dissipations, which are implemented by combining unitary processes with optical pumping of the electron spin that serves as an ancilla.
We show that this preparation is independent of the initial states of the nuclear spins.

\begin{figure}[tbp]
\includegraphics[width=0.47\textwidth]{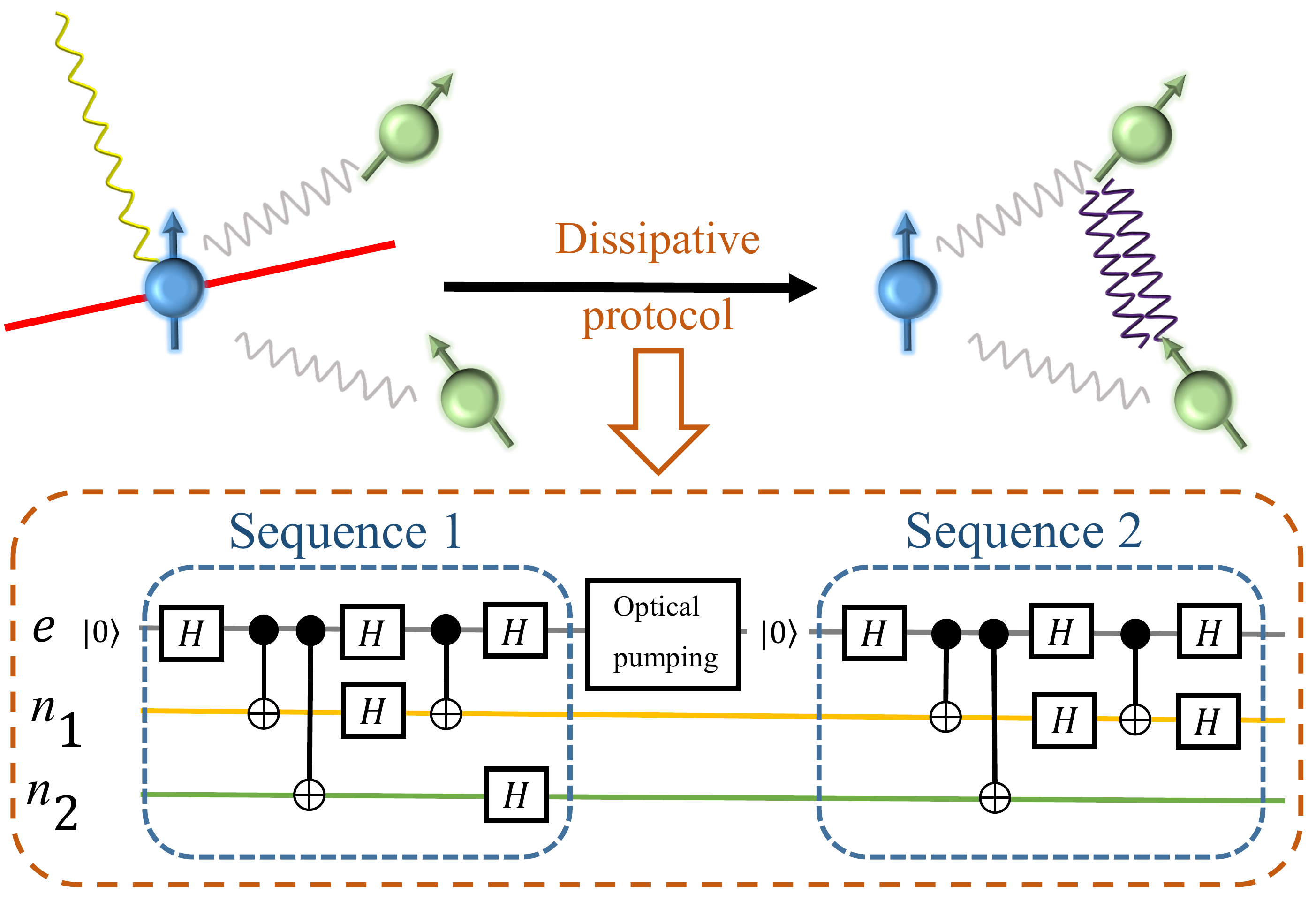}
\caption{An illustration of the dissipative protocol.
The electron spin (blue ball) has intrinsic hyperfine
interaction (grey wave) with surrounding $^{13}$C nuclear spins. By utilizing laser (red line) and microwave pulse (yellow wave), we can manipulate the electron and nuclear spins. Using the electron spin as an ancillary qubit (serving as the environment), we can implement the dissipative protocol to entangle deterministically the two nuclear spins. The below dashed
brown box shows the quantum circuit for the dissipative protocol, which involves only the Hadamard ($H$) and controlled-not gates. After the two sequences 1 and 2, the nuclear spins will be pumped into the GHZ state, irrespective of their initial state.}
\label{Illustration}
\end{figure}

\textit{The dissipative protocol for two qubits.}---We first introduce a simple and practical protocol for entangling two nuclear spins by dissipation in NV systems. We consider the NV electron spin as an ancilla (the environment) and the two most strongly coupled nuclear spins as the targeted system (the principal system). We denote them by  $e$, $n_1$, and $n_2$, respectively. In general, any dynamics of the total system (electron and nuclear spins) can be described by a unitary transformation $\rho_{en_1n_2}\mapsto U \rho_{en_1n_2} U^{\dagger}$, with $\rho_{en_1n_2}$ the joint density matrix of the whole system. Hence, the corresponding dynamics of the two nuclear spins reads $\rho_{n_1n_2}\mapsto \text{Tr}_e(U \rho_{en_1n_2} U^{\dagger})$. More conveniently, one may also describe the time evolution of the principle system in the operator-sum representation \cite{Nielsen2010quantum},
\begin{eqnarray}
\rho_{n_1n_2}\mapsto \mathcal{E}(\rho_{n_1n_2})\equiv \sum_k E_k \rho_{n_1n_2} E_k^{\dagger},
\end{eqnarray}
where $\mathcal{E}$ denotes the general quantum operation and its corresponding operators $\{E_k\}$ satisfy $\sum_k E_k^{\dagger}E_k=1$. Thus,  the task of dissipatively preparing the principle system into an entangled state reduces to   implementing appropriate sequences of dissipative maps that drive the system to the desired target state.  For our purpose, we use two simple unitary sequences together with optical pumping of the electron spin, as shown in Fig. \ref{Illustration}, to implement  two dissipative maps $\mathcal{E}_z$ and $\mathcal{E}_x$ in succession, which will drive the two nuclear spins into the target GHZ state irrespective of their initial state. In fact, direct calculations show that the operation elements $\{E^x_k\}$  for $\mathcal{E}_x$ (corresponding to sequence 2) reads: $E^x_0=\frac{1}{\sqrt{2}}(F_0+F_1)$ and $E^x_1=\frac{1}{\sqrt{2}}(F_0-F_1)$ with $F_0=\frac{1}{2}(1+X_{n_1}X_{n_2})$ and $F_1=\frac{1}{2}Z_{n_1}(1-X_{n_1}X_{n_2})$. Similarly, one can also obtain the operation elements $\{E^z_k\}$  for $\mathcal{E}_z$ (corresponding to sequence 1): $E^z_0=(H\otimes H)E^x_0$ and $E^z_1=(H\otimes H)E^x_1$ with $H$ denoting the Hadamard gate. It is then straightforward to check that $\mathcal{E}_z$ and $\mathcal{E}_x$ maps an arbitrary two-qubit state to the GHZ state.

To obtain a clearer idea on how this works, it is helpful to reexamine the two sequences in the Heisenberg picture. At the beginning of the protocol, we first polarize the electron spin to state $|0\rangle$, which is the eigenstate of the Pauli-Z matrix. The initial state for the two nuclear spin are arbitrary. Without loss of generality, we may assume that it is a pure state (our results will hold for mixed initial states as well since an arbitrary mixed state is just an ensemble of pure states).  For simplicity and conciseness, we use lower-case letters $x$, $y$, $z$ and $i$ to represent the single qubit eigenstates of three Pauli operators $X$, $Y$, $Z$ and identity operator $I$. We then apply a Hadamard gate $H_e$ on the electron spin, so as to obtain the state $x_ei_{n_1}i_{n_2}$. For this three-qubit state $x_ei_{n_1}i_{n_2}$, we apply two consequent controlled-not gates $C_{en_1}$ and $C_{en_2}$, with the electron spin being the control qubit, to evolve it to the state $x_ex_{n_1}x_{n_2}$.
Then we apply a controlled-not gate $C_{n_1e}$, where the first nuclear spin is the control qubit, to obtain the state $i_ex_{n_1}x_{n_2}$. Finally, we apply two Hadamard gates $H_{n_1}$ and $H_{n_2}$ on the two nuclear spins to rotate the state to $i_ez_{n_1}z_{n_2}$. This gives sequence $1$ in Fig. \ref{Illustration}. Here, we have replaced $C_{n_1e} $ with an equivalent sequence $H_eH_{n_1}C_{en_1}H_eH_{n_1}$, since in our NV system it is more convenient to implement the controlled-not gate $C_{en_1}$, rather than  $C_{n_1e} $. In the Heisenberg picture, the sequence 1 evolves the initial operator $Z_eI_{n_1}I_{n_2}$ to $I_eZ_{n_1}Z_{n_2}$. Thus, the sequence 1 will project the initial state of the two nuclear spins onto the $+1$ eigenspace of $Z_{n_1}Z_{n_2}$. Similarly, the sequence 2 will project the two nuclear spin state onto the $+1$ eigenspace of $X_{n_1}X_{n_2}$. In addition, since $Z_{n_1}Z_{n_2}$ commutes with $X_{n_1}X_{n_2}$, the sequence 1 (2) will keep the $+1$ eigenspace of $X_{n_1}X_{n_2}$ ($Z_{n_1}Z_{n_2}$) invariant. As a result, any two-qubit state going through sequences 1 and 2 will end up with the GHZ state, which is the only common eigenstate of both $Z_{n_1}Z_{n_2}$ and $X_{n_1}X_{n_2}$ with eigenvalue $+1$. Now, the physical meaning of $\mathcal{E}_z$ and $\mathcal{E}_x$ become apparent as well:  $\mathcal{E}_z$ pumps an arbitrary state to the $+1$ eigenspace of $Z_{n_1}Z_{n_2}$, whereas  $\mathcal{E}_x$ pumps  to the $+1$ eigenspace of $X_{n_1}X_{n_2}$, and the composition of them maps arbitrary two-qubit states to the GHZ state deterministically.

\begin{figure}[tbp]
\includegraphics[width=0.49\textwidth]{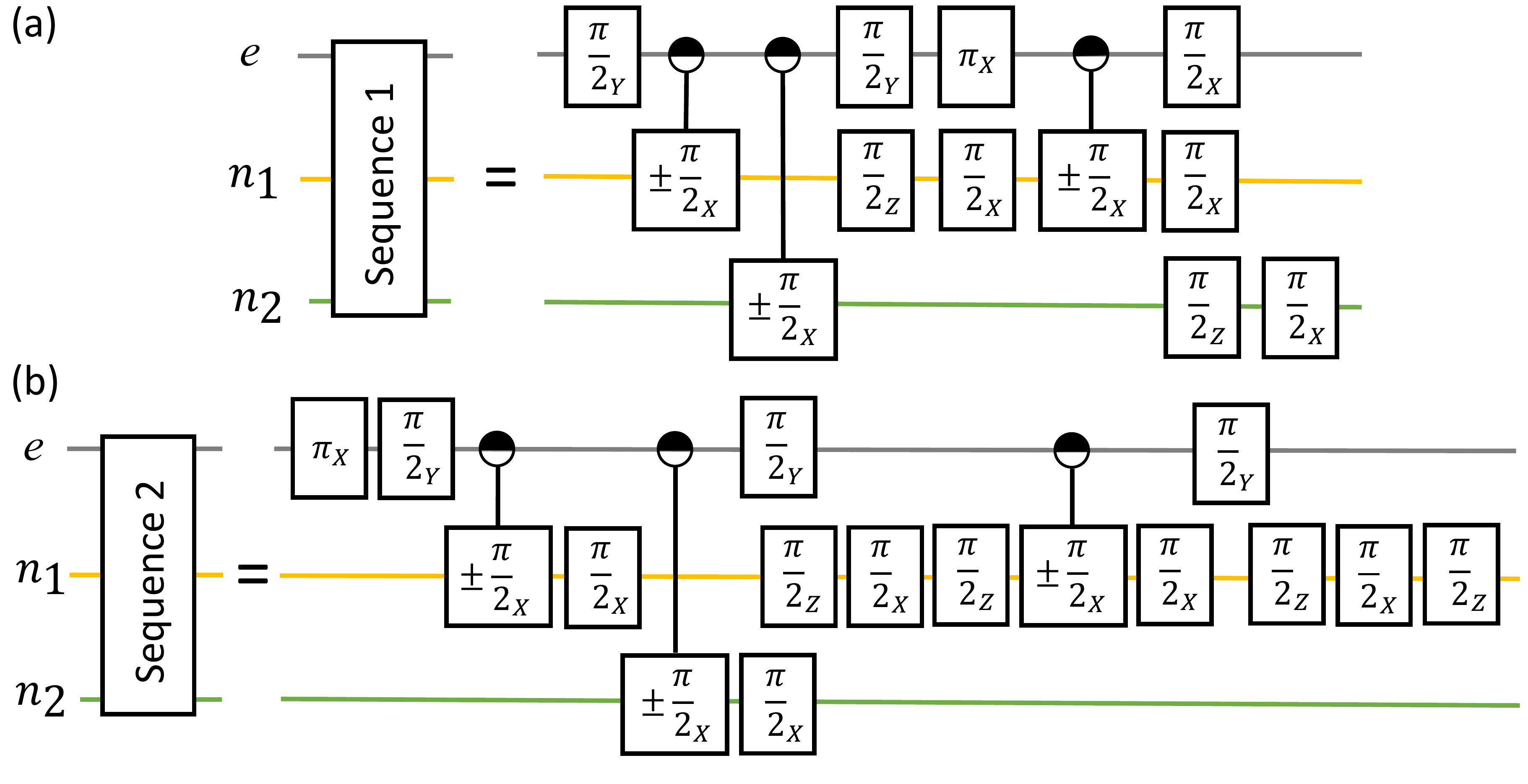}
\caption{The practical circuit diagrams for implementing the two sequences 1 (a) and 2 (b) in the dissipative protocol in our experiment. }
\label{S1S2}
\end{figure}

\textit{Experimental setup.}---We perform the experiment under a cryogenic temperature (about 7K) on a type-IIa chemical-vapor-deposition (CVD) synthetic diamond sample with the natural abundance of $^{13}$C (about 1.1$\%$). As shown in Fig. \ref{NVIllustration}(b), the NV electron spin can be optically polarized and read out by the single-shot technique \cite{Robledo2011High} with red laser beams of wavelength about $637.2$nm (the polarization fidelity is about $99\%$ and the average fidelity for the read out is about $76.5\%$ in our experiment).  To improve the precision of the experimental result, we repeat the single-shot readout $5000$ times.  Microwave is used in the experiment to rotate the electron spin and a magnetic filed about $493$ Gauss along NV axis is also applied.
We use the dynamical decoupling  method \cite{Hanson2008Coherent,Ryan2010Robust} to decouple the nuclear spins from the electron spin and obtain their hyperfine interaction parameters ~\cite{CPMG}, which are used to design the desired quantum gates.
 We apply a XY-8 type Carr-Purcell-Meiboom-Gill (CPMG) sequence on the NV center to sense  different weakly coupled $^{13}$C nuclear spins [see FIG.~\ref{NVIllustration}(c)], and use the microwave $\pi$ pulse to flip the electron spin. The nuclear spins precess around different axes depending on the state of the electron spin.  When the electron spin is in the state $|m_s = 0\rangle$, all nuclear spins precess with Larmor frequence $\bm{\omega}_L$ around the axis parallel to the magnetic field along the nitrogen-vacancy axis. Whereas when the electron spin is in the state $|m_s = -1\rangle$, each nuclear spin will precess around a certain axis $\bm{\omega} = \bm{\omega}_L + \bm{\omega}_h$, where $\bm{\omega}_h$ depends on the relative position of the nuclear spin with respect to the electron spin.
Because of the multiple $\pi$ pulses in the CPMG sequence, different $^{13}$C nuclear spins  will precess around different axes associated with their hyperfine interactions, hence they can be decoupled and isolated through the sequence. We utilize an adaptive method developed in our previous work \cite{Adaptive}  to measure more efficiently  the hyperfine interaction parameters for all the nearby nuclear spins around the NV center.



\begin{figure}[tbp]
\includegraphics[width=0.47\textwidth]{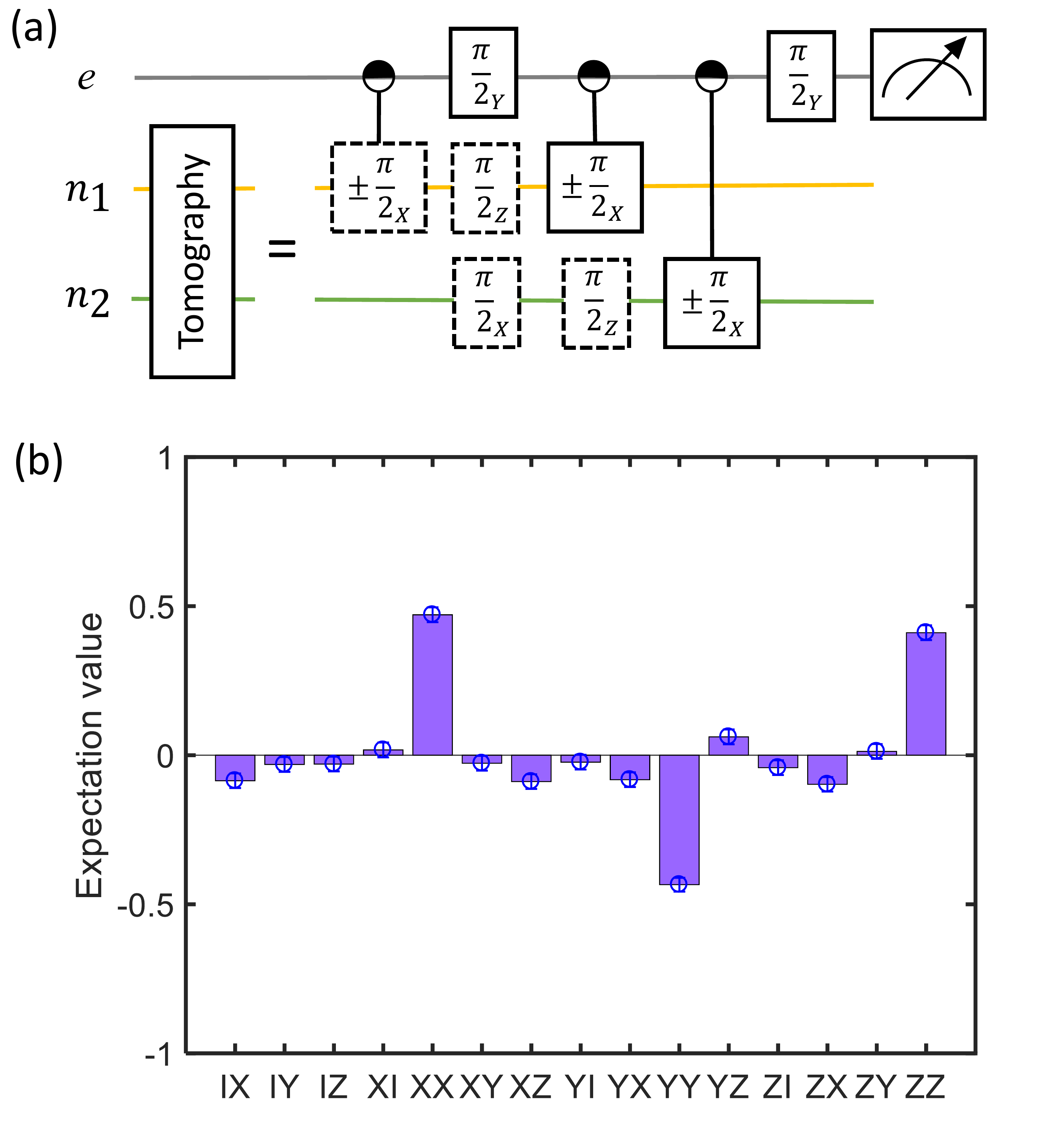}
\caption{Two-qubit tomography of the nuclear spins entangled by the dissipative protocol.
(a)The gate circuits used in our experiment for tomography, which map the expectation values of the nuclear spins onto the electron spin \cite{taminiau2014universal}.  Here, the electron spin is prepared to be in the state $|0\rangle$ before the tomography and the gates with dashed rectangle are optional basis rotations.
(b)The experimental tomography result for the dissipatively generated entangled nuclear spin state.  }
\label{Tomography}
\end{figure}

By tuning the two parameters $\tau$ and $N$ in the XY-8 CPMG sequence, we can realize the electron-controlled-nuclear not gate ($C_{en_1}$ and $C_{en_2}$), and single nuclear $\pi/2$ rotations around $x$ or $z$ axis~\cite{taminiau2014universal}. These gates provide an convenient implementation of the sequences 1 and 2 desired  for our dissipative protocol.  Based on this, the practical circuits for the implementation of the sequences 1 and 2 in our experiment are shown in Fig. ~\ref{S1S2}. To characterize the dissipatively prepared entangled state of the nuclear spins, we perform a tomography process to obtain the density state and estimate the fidelity between the obtained state and the ideal GHZ state. This is accomplished by mapping the the expectation values of the Pauli operators for the nuclear spins onto the electron spin and reading out the electron \cite{taminiau2014universal}.  The circuits used in our experiment for the mapping  is shown in Fig.~\ref{Tomography}(a). We mention that some phase compensations might be necessary during the tomography process due to the cross-talk effect between nuclear spins and imperfections of the unitary gates used in the circuit (see the Appendices for details).



\textit{Experimental results.}---To start our experiment, we first pumping the electron spin to state $|0\rangle$ and carry out no operations on the target two nuclear spins. Thus, the initial state of the two nuclear spins is a complete mixed state. We then perform the sequence 1 and 2 to implement the dissipative protocol and evolve the nuclear state to the desired GHZ state. After this, we carry out the tomography process to obtain the density state for the nuclear spins and estimate its fidelity. Our tomography result is plotted in Fig~\ref{Tomography}(b), which shows the raw data renormalized by the electron Rabi contrast between states $|0\rangle$ and $|-1\rangle$. We use the maximum likelihood method to estimate the density matrix of the generated state and calculate its fidelity~\cite{likelihood}. For the state plotted in Fig~\ref{Tomography}(b), the estimated fidelity is about $0.579\pm 0.011$, which is larger than $1/2$, indicating that the two nuclear spins are indeed entangled.



\begin{figure}[tbp]
\includegraphics[width=0.49\textwidth]{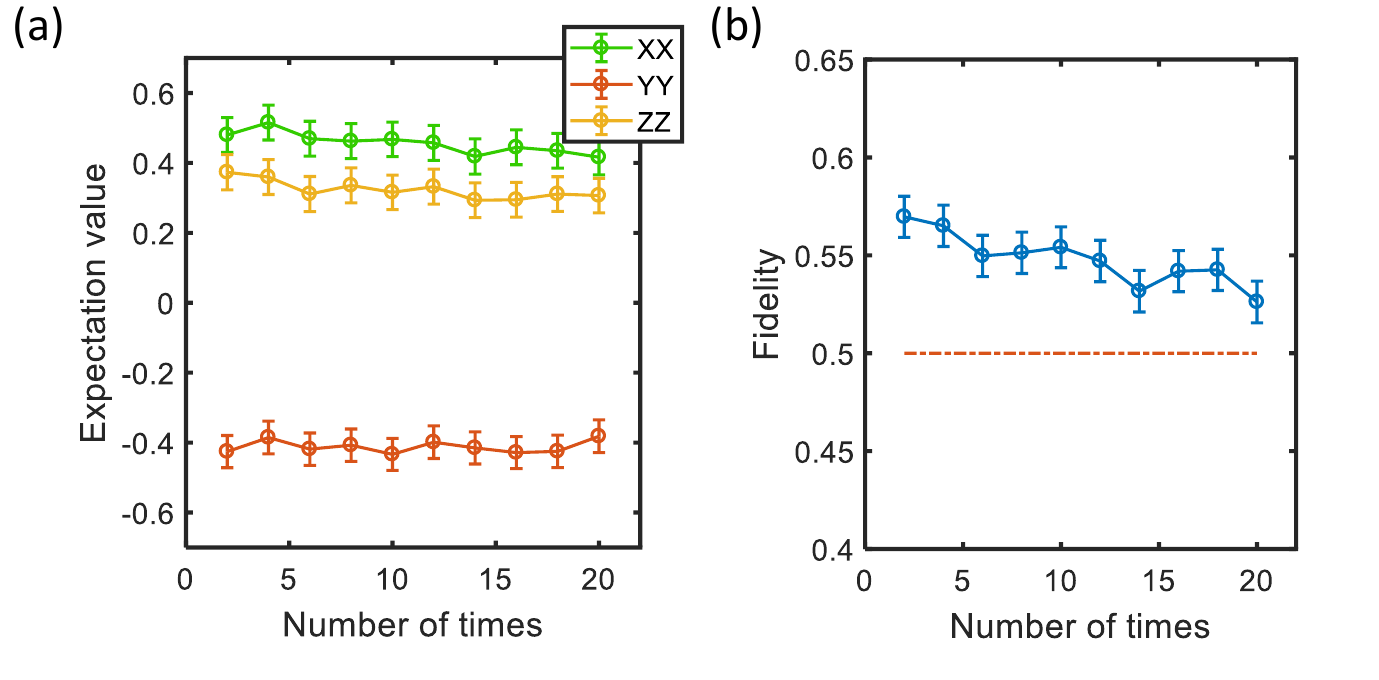}
\caption{(a) Correlation $\langle X_{n_1}X_{n_2} \rangle$, $\langle Y_{n_1}Y_{n_2} \rangle$ and $\langle Z_{n_1}Z_{n_2}\rangle$
for the nuclear spins as a function of number of repeated times in applying the dissipative protocol.  (b) The corresponding fidelity. }
\label{vsN}
\end{figure}

To further illustrate that the dissipative protocol is independent of the initial state of the nuclear spins and the sequence 1 and 2 would stabilize the generated entanglement against environment noises, we successively apply the sequences 1 and 2 for a number of times. In this case, the measured correlation functions and estimated fidelity are shown in Fig. \ref{vsN}. Here, for simplicity we estimate the fidelity by the formula $F=1/2-\langle \hat{W}\rangle$, with  $\hat{W}=\frac{1}{4}(1-XX+YY-ZZ)$ being the witness for the GHZ state ~\cite{BellF}. From this figure, it is evident that the correlations $\langle X_{n_1}X_{n_2} \rangle$, $\langle Y_{n_1}Y_{n_2} \rangle $, and $\langle Z_{n_1}Z_{n_2} \rangle$, and the estimated fidelity remain stable as we repeatedly apply the sequences 1 and 2. The fidelity is always larger than $1/2$, showing that the nuclear spins remain entangled.

We mention that for low-temperature NV system, the previous work has achieved entanglement of two nuclear spins by non-destructive measurements with a fidelity $0.824(7)$ after calibration and correction of two-qubit readout error \cite{cramer2016repeated}. In our experiment, the measured fidelity is about $0.579$ for the raw data, which is notably smaller than the previously reported result. A two-qubit readout calibration following Ref. \cite{cramer2016repeated} will increase the fidelity to $0.708\pm 0.011$. From the results shown in Fig. \ref{vsN}, the correlations and fidelity for the nuclear spins remain stable after multiple running of the dissipative protocol, implying that the error induced by non-perfect optical pumping of the electron spin is negligible. The major factor that brings down the fidelity in our experiment is the strong cross-talk effect between nuclear spins. For the NV sample used in our experiment, we find that there is a $^{13}$C with $A_{zz} = -1.2969(2)$MHz coupled to the electron spin, whose CPMG signal has a very broad range in time (see appendix). This makes it difficult in choosing optimal gate parameters $\tau$ and $N$ for designing elementary gates and leads to large imperfections for implementing sequence 1 and 2. In fact, here we have chosen two nuclear spins with highest  polarization fidelity among nearby  nuclear spins weakly coupled to the electron spin. Yet, their polarization fidelity are only $0.896 \pm 0.025$ and $0.829 \pm 0.019$ (see appendix), which reflects the large imperfections in the elementary gates and indicates a very strong cross-talk effect between the nuclear spins. As a result, the fidelity of the dissipatively prepared entangled state is substantially reduced. We expect that choosing a better NV sample will significantly improve the fidelity.

\textit{Summary.} ---We have demonstrated a dissipative protocol to prepare two nuclear spins, a good candidate of memory qubits in a solid-state sample, into the GHZ state, through their coupling to the electron spin state and controlled dissipation of the system with optical pumping. The produced entangled state of the nuclear spins is independent of their initial mixed state and stays alive as an steady state of the system under repeated application of the engineered dissipation. Our demonstration of preparing entangled states through engineered dissipation provides a new tool for quantum information processing with solid-state qubits.

%
%
%
%

\begin{acknowledgments}
This work was supported by the Frontier Science Center for Quantum Information of the Ministry of Education of China, Tsinghua University Initiative Scientific Research Program, and the National key Research and Development Program of China (2016YFA0301902).
\end{acknowledgments}

\begin{appendix}
\section{The diamond sample}

The diamond sample (Element Six) used in this work has a natural abundance of $^{13}$C atoms (about $1.1\%$) and $\langle100\rangle$ crystal orientation. We fabricate solid immersion lens (SIL) and waveguide on the surface of the sample.
In the surface fabrication procedure, we make some markers by a focused ion beam (FIB, FEI, Helios Nanolab 660) on the sample surface and then search NV centers which are about 5$-$15 $\mu$m beneath the surface with a room temperature optical setup. We record the relative position between a NV center and a nearby marker to localize a NV. After that we use FIB to make a hemisphere structure around the NV center. SILs on the surface enhance the photoluminescence intensity of NV for about 7 times, which improve the collection efficiency of fluorescence photon and makes single-shot readout possible under low temperature. In our experiment, the photoluminescence rate of NV center excited by 532nm laser at saturation power is about 320kcts$\cdot s^{-1}$.

There is a gold waveguide with thickness of about 2 $\mu$m around the SIL, fabricated by lithography (SUSSMicroTec, SUSS MA/BA6) used to deliver the microwave signal. We also build some electrodes on the diamond surface near SILs to tune the strain by applying a DC electric field ~\cite{Chu2015}.

\begin{figure}[tbp]
	\centering
	\includegraphics[width=3.3 in]{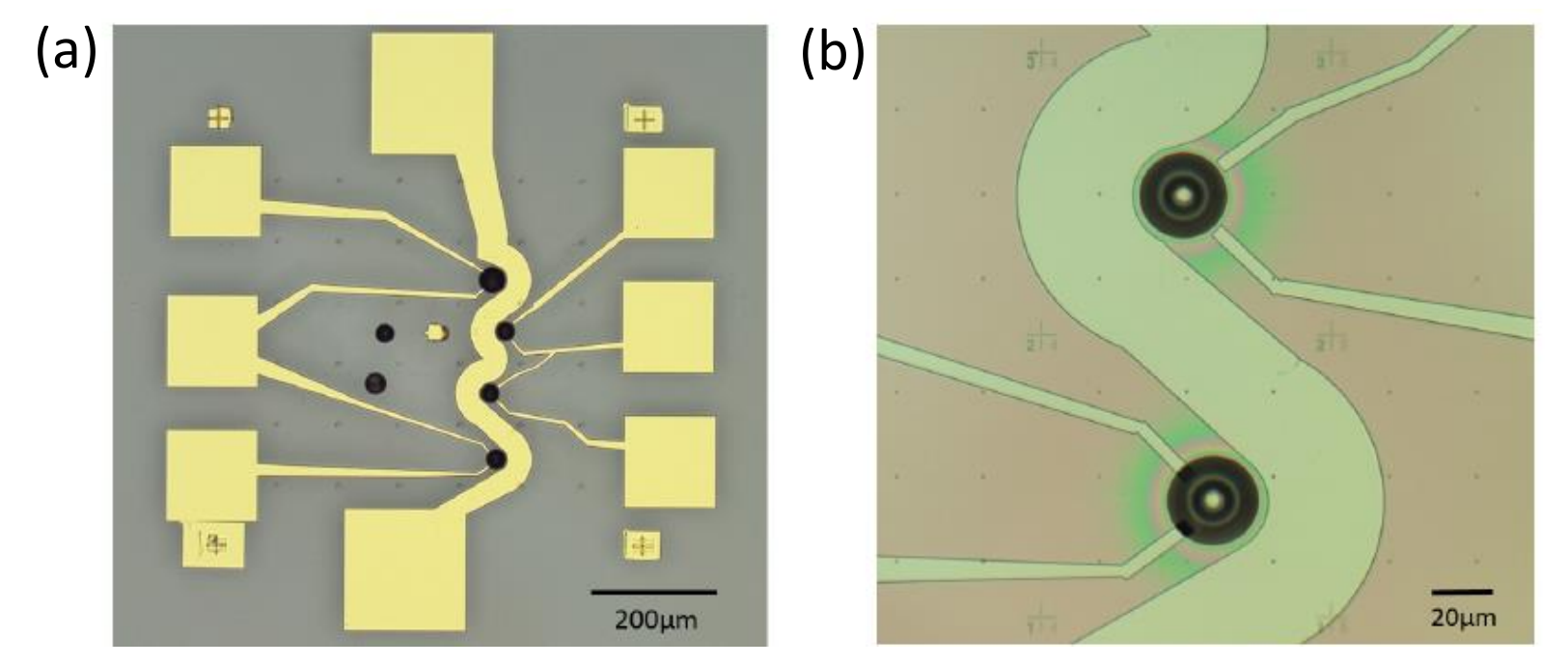}
	\caption{Example photos of the diamond sample surface. (a) A photo of the diamond sample surface, including SILs (black circles), waveguide and electrodes (yellow pattern).
(b) A zooming-in image around SILs.}
	\label{Sample}
\end{figure}

\section{Experimental Setup}
Our experiment is performed in low temperature about 7K in a commercial cryostation (Montana Instruments, Nanoscale Workstation). A high-NA objective lens with NA = 0.9 (Zeiss) is inside the cryostation. A heater keeps the lens at temperature about 305K to protect it from low temperature. The diamond sample is fixed on a three-dimensional positioner (Attocube, ANC 350 controller) inside the low temperature chamber, which is able to adjust the relative position between the sample and laser beams.

Three laser beams are utilized in our experiments. A 532nm laser (Coherent) is used to ionize the charge state of the NV center. Two red lasers with wavelength about 637.2nm (New focus) are used to initialize and readout states of the NV electron spin. Each laser is controlled by  an acoustic optical modulator (AOM, Isomet 1250C-848), which improves the laser on-off ratio to $10^5 : 1$. Three laser beams pass through two dichroic mirrors (Semrock) and an optical 4f system, consisting of a two-dimensional galvo scanning mirror (Thorlabs, GVS212) and two lenses. The fluorescence emitted from the NV is coupled to a multi-mode fiber and detected by a single photon detector (Excelitas, SPCM-AQRH-14-FC).

Microwave signal is generated by a microwave source (Keysight N5181B). An arbitrary waveform generator (AWG, Tektronix 5014C) generates two signals with the same frequency of 100MHz, same amplitude, same time resolution of 1ns and a $\pi/2$ relative phase difference. An IQ mixer (Marki Microwave IQ-1545LMP) combines these two signals with the signal generated from microwave source to create a new signal with frequency corresponding to the electron transition $|m_s=0\rangle \leftrightarrow |m_s=-1\rangle$. Similarly, a signal with frequency corresponding to the electron transition $|m_s=0\rangle \leftrightarrow |m_s=+1\rangle$ is also generated.
Two amplifiers (Minicircuits ZHL-30W-252-S+ and a home-made amplifier) are used for amplifying these two signals.

A permanent magnet is set on a three-axis stage (Thorlabs, MT3-Z8) to apply an external field about 493 Gauss along the NV symmetry axis (regarded as $z$ axis). The three-axis stage helps to optimize the magnet field around the NV center by finding out a position with minimum value of the magnet field component in the $x-y$ plane.

\section{Experimental procedure}
Before performing our dissipative protocol, we need to make sure that the charge state of the NV center is NV$^{-}$ and polarize the nitrogen nuclear spin. By using the technique reported in Ref ~\cite{cramer2016repeated,Robledo2011High}, we check the charge state and the nitrogen polarization status during the experimental procedure.

\begin{figure}[tbp]
	\centering
	\includegraphics[width=3.3 in]{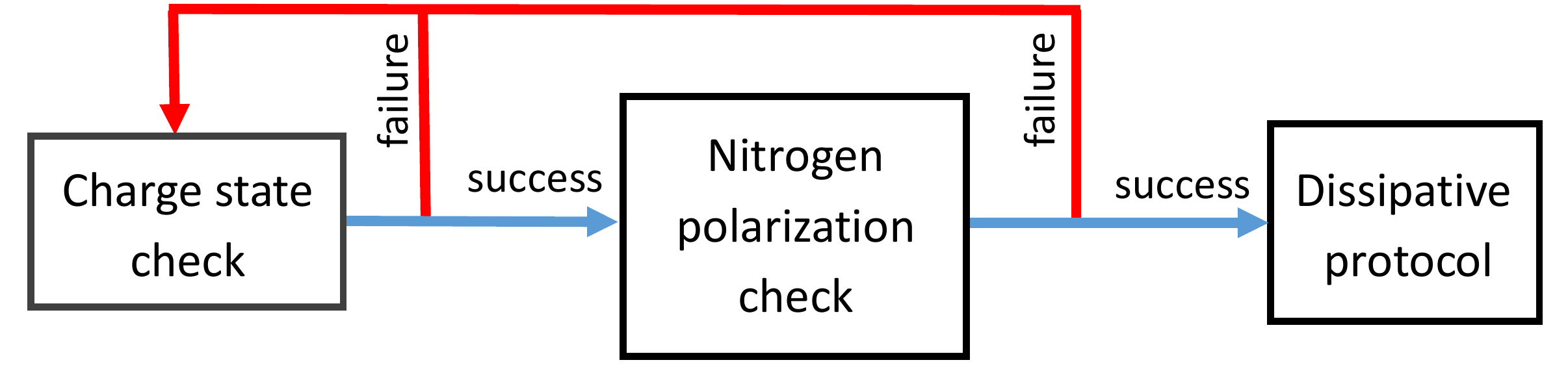}
	\caption{An illustration of the experimental procedure.}
	\label{ExpProcedure}
\end{figure}

We first check the charge state of the NV center by turning on two red lasers at the same time. A threshold of photon numbers in a single-shot readout can tell the difference between NV$^{-}$ and NV$^0$. After checking the charge state, we polarize the nitrogen nuclear spin to $|m_N = -1\rangle$ state under low temperature~\cite{Robledo2011High}. We regard that the polarization is successful if at least one photon is detected in a single-shot readout. The experimental procedure is shown in Fig. \ref{ExpProcedure}.

\section{hyperfine interaction parameters of $^{13}$C nuclear spins}
To design quantum gates, we first need to find out the hyperfine interaction parameters between $^{13}$C nuclear spins and the NV electron spin. We use the dynamical decoupling technique  to accomplish this and to manipulate the nuclear spins\cite{CPMG}. A relatively strong magnetic field $B$ along the NV axis is applied in such operations. With the rotating wave and secular approximations,  the effective Hamiltonian for the system consisting of the NV electron and a single coupled nuclear spin reads:

\begin{eqnarray}
\bm{\hat{H}_{eff}} &=& \omega_L\bm{\hat{I}_z}+A_{zz}\bm{\hat{S}_z}\bm{\hat{I}_z}+A_{zx}\bm{\hat{S}_z}\bm{\hat{I}_x}\nonumber \\
&=&|0\rangle\langle0|\bm{\hat{H}_0}+|-1\rangle\langle-1|\bm{\hat{H}_{-1}}+|+1\rangle\langle+1|\bm{\hat{H}_{+1}}, \nonumber
\end{eqnarray}
with
\begin{eqnarray}
\bm{\hat{H}_0} = \omega_L\bm{\hat{I}_z}, \;
\bm{\hat{H}_{\pm 1}} = (\omega_L\pm A_{zz})\bm{\hat{I}_z}+A_{zx}\bm{\hat{I}_x}, \label{Hams}
\end{eqnarray}
 where $\bm{\hat{H}_0}$ and $\bm{\hat{H}_{\pm 1}}$ are Hamiltonians when the electron spin is at $|0\rangle$ and $|\pm 1\rangle$ states respectively,
  $\omega_L=\gamma_nB_z$ with $\gamma_n$ being the gyromagnetic ratio of the nuclear spin,
 $B_z$ is the $z$ component of the magnetic field,
$A_{zz}$ and $A_{zx}$ are hyperfine interaction parameters between the electron and the nuclear spins, 
$\bm{\hat{I}_{x,y,z}}$ are spin-1/2 Pauli operators for nuclear spins, and $\bm{\hat{S}_z}$ is the Pauli-Z operator for the electron spin. 

From Eq. (\ref{Hams}), we see that when the electron spin is at $|0\rangle$ state, the nuclear spin will precess around $z$ axis with Larmor frequency $\omega_L$. When the electron spin is at $|\pm1\rangle$ states, the precession axis will deviate from $z$ axis slightly and the precession frequency will change because of the hyperfine interaction. Based on this, we can calculate hyperfine parameters $A_{zz}$ and $A_{zx}$ of a single nuclear spin from free precession frequencies by using the following equation:
\begin{equation}
f_{\pm}=1/2\pi\sqrt{{A_{zx}}^2+{(A_{zz}\pm\omega_n)}^2}, \label{freq}
\end{equation}
where $f_{\pm}$ are the free precession frequencies of the nuclear spin when the electron spin is at $|\pm 1\rangle$ states, respectively.

We need to know the rough values of $A_{zz}$ and $A_{zx}$ to polarize a single nuclear spin through quantum gates. By fitting the experimental CPMG signal with a simulation of estimated hyperfine parameters, we can obtain these two parameters roughly~\cite{CPMG}.
After that we can perform single nuclear Ramsey-type free precession to calibrate these two parameters~\cite{taminiau2014universal}. We polarize the nitrogen and a single $^{13}$C nuclear spin, then measure the precession frequency when the  electron spin is at $|\pm1\rangle$ states. By using Eq. (\ref{freq}) and the known magnetic field along the NV axis, we can calculate the hyperfine parameters more precisely.

 In our experiment, we utilize an adaptive method to measure the hyperfine parameters more efficiently~\cite{Adaptive}. The adaptive method includes a sequence of Ramsey-type experiments, which gradually narrows the frequency estimation range.

\begin{figure}[tbp]
	\centering
	\includegraphics[width=0.49\textwidth]{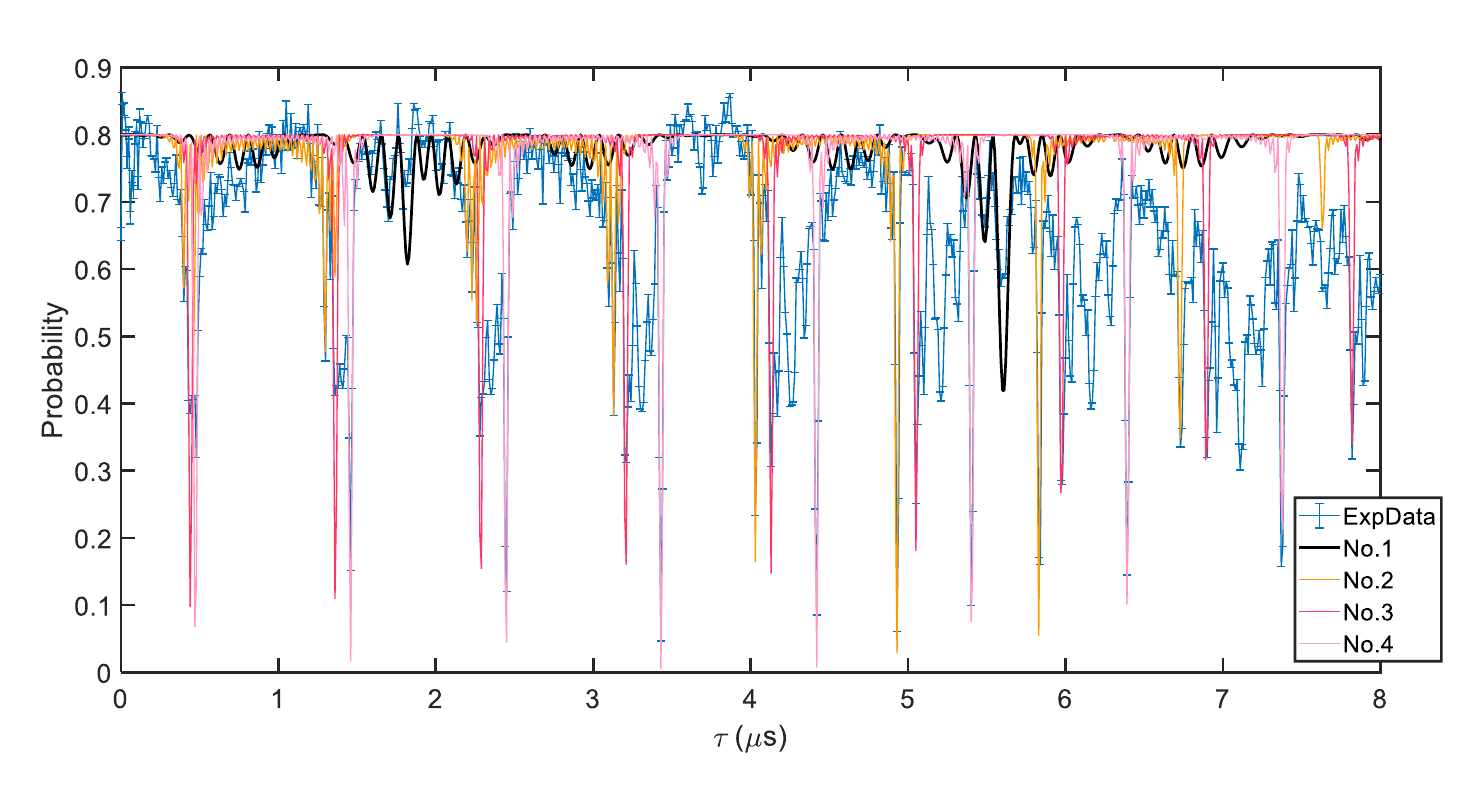}
	\caption{The experimental CPMG figure and the simulation signals of nuclear spins with hyperfine parameters in TABLE. ~\ref{HypPara}. The blue line represents the experimental data and other lines with different colours represent a single nuclear spin.}
	\label{CPMGexp}
\end{figure}

The hyperfine parameters of nuclear spins around the NV center we have detected are shown in TABLE. ~\ref{HypPara}. The experimental CPMG figure and the simulation signals of nuclear spins with hyperfine parameters in TABLE. ~\ref{HypPara} are shown in FIG. \ref{CPMGexp}. We see that the signal of No.$1$ nuclear spin (black line in FIG. \ref{CPMGexp}) has a broader range in time compared with other weakly coupled nuclear spins. It also has relative strong oscillation in some ranges of time, which influences the choice of gate parameters of other nuclear spins.

\begin{table}

\caption{Hyperfine interaction parameters of $^{13}$C nuclear spins around the NV center.}

\label{tab:1}

\begin{tabular}{c|c|c}

\hline\noalign{\smallskip}

Number & $A_{zz}$ (kHz) & $A_{zx}$ (kHz)  \\

\noalign{\smallskip}\hline\noalign{\smallskip}

1 & -1296.9(2) & 180(1) \\
2 & 50.16(7)  & 101.6(4) \\
3 & 30.62(5)   & 43.0(7) \\
4 & -41.20(7)  & 52.3(7) \\

\noalign{\smallskip}\hline

\end{tabular}
\label{HypPara}
\end{table}

\section{Polarization of $^{13}$C nuclear spin}
\begin{figure}[tbp]
	\centering
	\includegraphics[width= 2.5in]{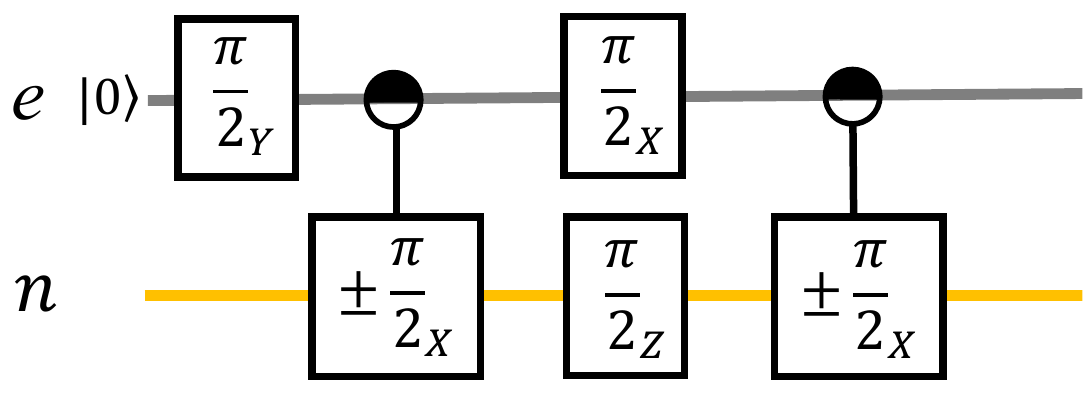}
	\caption{Quantum gate circuit of $^{13}$C nuclear spin polarization~\cite{taminiau2014universal}.}
	\label{CPolarize}
\end{figure}

When performing the nuclear free precession to measure the hyperfine interaction parameters, we need to polarize the single $^{13}$C nuclear spin. The gate circuit of polarization is shown in Fig. \ref{CPolarize}.

We calculate the polarization fidelity of some weakly coupled $^{13}$C nuclear spins near the NV center and choose two nuclear spins with the highest fidelity (No.2 and No.4 nuclear spin in the TABLE. ~\ref{HypPara}.) to operate our dissipative protocol.
By renormalizing the data of the nuclear free precession with the contrast of electron Rabi oscillation, we are able to estimate the fidelity of the $^{13}$C nuclear spin. We fit the data with cosine function, and the visibility of the oscillation is

\begin{equation}
V=\frac{P_{max}-P_{min}}{P_{max}+P_{min}},
\end{equation}

where $P_{max}$ and $P_{min}$ are the maximum and minimum values of the fit function.

Then the estimated fidelity is
\begin{equation}
F = \frac{V+1}{2}.
\end{equation}

\begin{figure}[tbp]
	\centering
	\includegraphics[width= 3.6in]{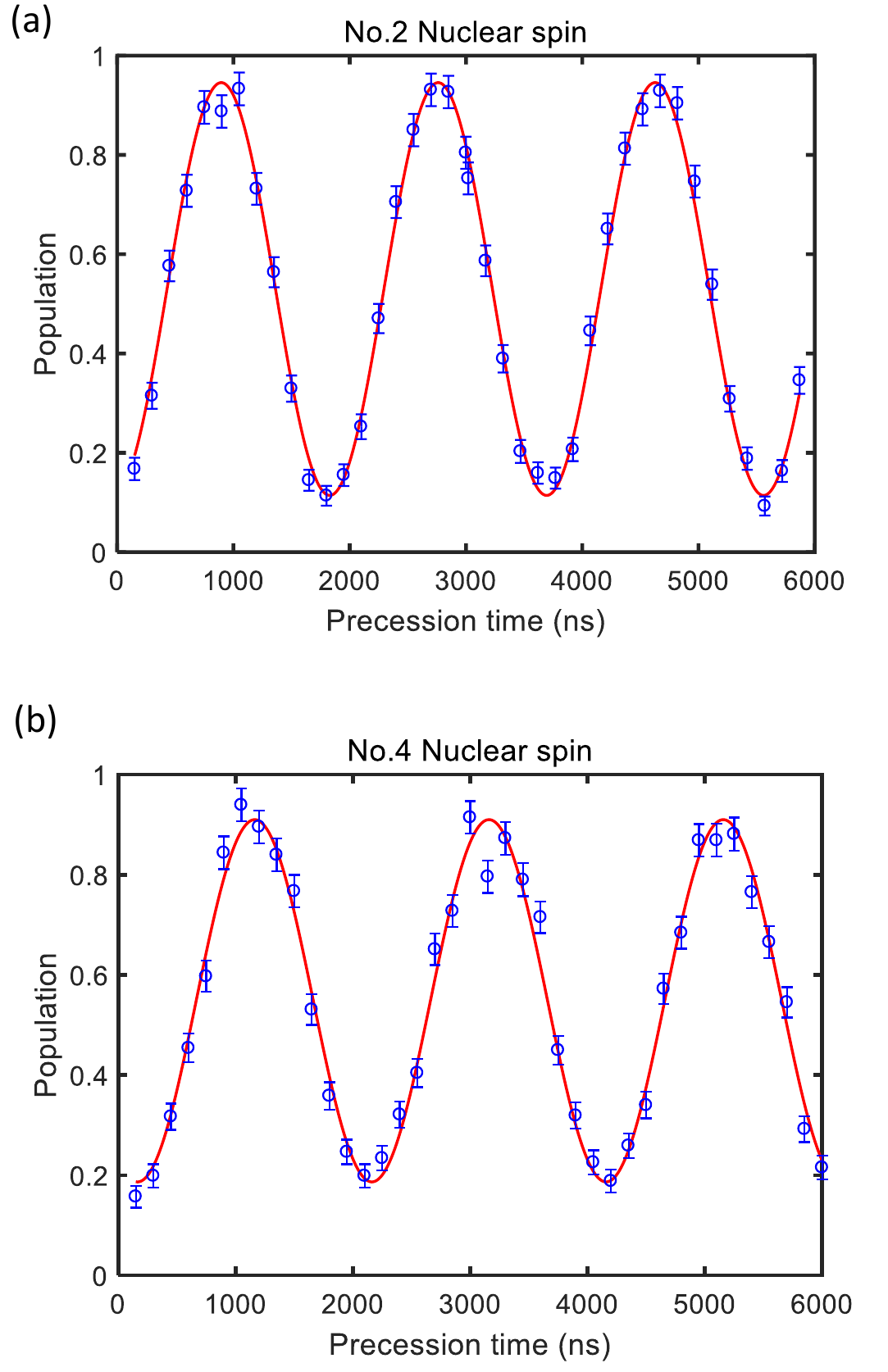}
	\caption{$^{13}$C nuclear free precession oscillation. The electron spin is at $|0\rangle$ state and the measurement basis is nuclear $Y$ basis. (a) The fidelity of No.2 nuclear spin is $0.896\pm 0.025$. (b) The fidelity of No.4 nuclear spin is $0.829\pm 0.019$.}
	\label{CCfidelity}
\end{figure}

We show the population on state $(|\uparrow\rangle+i|\downarrow\rangle)/\sqrt{2}$ in single nuclear $Y$ basis varies with free precession time when the electron spin is at $|0\rangle$ state in Fig. \ref{CCfidelity}. The estimated fidelity for the two nuclear spins are $0.896\pm 0.025$ and $0.829\pm 0.019$.

The gate parameters we used for our dissipative protocol for two nuclear spins are shown in TABLE. \ref{GatePara}.

\begin{table}

\caption{Quantum gate parameters for two nuclear spins\\(at magnetic field $B_z = 492.65$ Gauss)}

\label{tab:1}

\begin{tabular}{c|c|c|c}

\hline\noalign{\smallskip}

Nuclear spins &   Quantum gate   & $\tau (ns)$  &  $N$  \\

\noalign{\smallskip}\hline\noalign{\smallskip}

No.2 & $R^{\pi /2}_{\pm x}$ & 4915 & 16 \\
  & $R^{\pi /2}_{z}$  & 37 & 4 \\
  & $R^{\pi /2}_{x}$  & 6260 & 10\\
  \noalign{\smallskip}\hline\noalign{\smallskip}
No.4 & $R^{\pi /2}_{\pm x}$ & 4411  & 16 \\
  & $R^{\pi /2}_{z}$ & 36 & 4 \\
  & $R^{\pi /2}_{x}$ & 5886 & 22 \\

\noalign{\smallskip}\hline

\end{tabular}
\label{GatePara}
\end{table}

\section{Error bars of raw data}
 We suppose that measurement result of the electron state ($|0\rangle$ or $|-1\rangle$) in our experiment obeys Bernoulli distribution. Then we calculate the standard deviation of a Monte Carlo simulation repeated $10^4$ times to obtain the error bars of raw data:

\begin{equation}
\sigma = 2f\sqrt{\frac{P_0(1-P_0)}{N}},
\end{equation}
where $P_0$ represents the raw data without electron Rabi normalization, $N$ is the repeating times of the single-shot readout and $f$ is a normalized factor defined by

\begin{equation}
f =1/(P_{Rmax}-P_{Rmin}),
\end{equation}
where $P_{Rmax}$ and $P_{Rmin}$ are the maximum and minimum values of electron Rabi oscillation.

\section{Simulation of tomography}
For the GHZ state $(|00\rangle+|11\rangle)/\sqrt{2}$, only four tomography basises have expectation value oscillation varying with separation time between state generation and tomography measurement. According to our simulation, we should compensate some phase during the tomography measurement.

Cross-talk between nuclear spins is considered to be the main reason of the phase compensation. It may cause some unwanted operations in tomography measurement. By tuning gate parameters $\tau$ and $N$ or compensate accumulated phase properly in the tomography sequence, we can reduce the cross-talk effect~\cite{DFS}.

\begin{figure}[tbp]
	\centering
	\includegraphics[width= 0.49\textwidth]{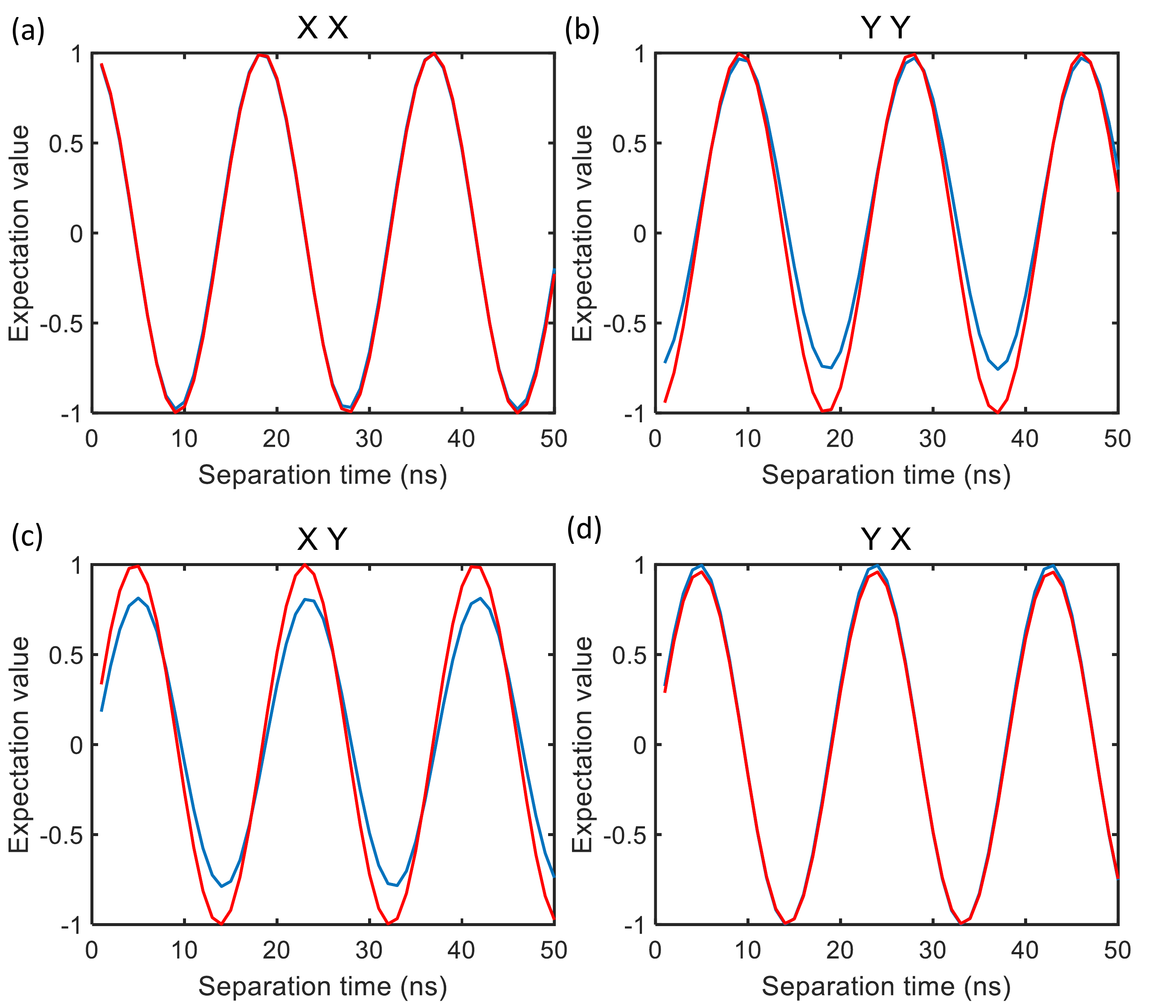}
	\caption{Simulation of the tomography measurement of $XX$, $YY$, $XY$ and $YX$ basises. The red line represents the simulation of ideal quantum gates, and the blue line represents the simulation with cross-talk effect.}
	\label{TomoSim}
\end{figure}

The simulation results of tomography measurement are shown in Fig. \ref{TomoSim}. After phase compensation, the simulation with cross-talk effect has almost no phase difference with the simulation consisting of ideal quantum gates in oscillation. Simulation result guarantees that the tomography measurement is correct at arbitrary separation time.

We see that the contrasts of the oscillation with cross-talk effect decrease slightly in some tomography basises.
The decrease also affects the estimated fidelity of the final generated state.

\section{Estimated fidelity of the generated state}
From the result of the tomography, we calculate the estimated density matrix of the generated state by a maximum likelihood method~\cite{likelihood}. The density matrix reads $\rho=\rho_R+i\rho_I$:
\begin{eqnarray}
\rho_R = \left[
 \begin{array}{cccc}
0.3706   & -0.0033  & -0.0263  &  0.2260 \\
-0.0033  &  0.1502  & 0.0096  &  0.0177 \\
-0.0263 & 0.0096   & 0.1446  &  0.0462 \\
0.2260   & 0.0177   & 0.0462  &  0.3346 
\end{array}
\right ] \nonumber
\\
\rho_I = \left[
 \begin{array}{cccc}
 0.0000  & 0.0109 & 0.0214 &  - 0.0273\\
 - 0.0109 & 0.0000  & - 0.0136 &   - 0.0095\\
- 0.0214 & 0.0136  & 0.0000 &   0.0047\\
0.0273  &  0.0095  & - 0.0047 &  0.0000
\end{array}
\right ] \nonumber
\end{eqnarray}

By comparing the estimated density matrix $D$ with the ideal density matrix of GHZ state $(|00\rangle + |11\rangle)/\sqrt{2}$, we obtain the estimated fidelity $F = 0.579 \pm 0.011$.

\end{appendix}

\bibliographystyle{apsrev4-1-title}
\bibliography{EntanglingNSRefs}

\end{document}